\documentclass[twocolumn,english,superscriptaddress,citeautoscript,showpacs,preprintnumbers,amsmath,amssymb,prb,floatfix,footinbib]{revtex4}
\usepackage[utf8x]{inputenc}
\usepackage[scaled=2]{helvet}
\usepackage[T1]{fontenc}
\usepackage[utf8]{luainputenc}
\usepackage{multirow}
\usepackage{textcomp}
\usepackage{amsmath}
\usepackage{amssymb}
\usepackage{graphicx}
\usepackage{subscript}
\usepackage{epstopdf}

\makeatletter

\newcommand{\lyxmathsym}[1]{\ifmmode\begingroup\def\b@ld{bold}
  \text{\ifx\math@version\b@ld\bfseries\fi#1}\endgroup\else#1\fi}

\providecommand{\tabularnewline}{\\}

\@ifundefined{textcolor}{}
{%
 \definecolor{BLACK}{gray}{0}
 \definecolor{WHITE}{gray}{1}
 \definecolor{RED}{rgb}{1,0,0}
 \definecolor{GREEN}{rgb}{0,1,0}
 \definecolor{BLUE}{rgb}{0,0,1}
 \definecolor{CYAN}{cmyk}{1,0,0,0}
 \definecolor{MAGENTA}{cmyk}{0,1,0,0}
 \definecolor{YELLOW}{cmyk}{0,0,1,0}
}

\usepackage[scaled=2]{helvet}\usepackage{amstext}

\makeatletter
\@ifundefined{textcolor}{}{\definecolor{BLACK}{gray}{0}\definecolor{WHITE}{gray}{1}\definecolor{RED}{rgb}{1,0,0}\definecolor{GREEN}{rgb}{0,1,0}\definecolor{BLUE}{rgb}{0,0,1}\definecolor{CYAN}{cmyk}{1,0,0,0}\definecolor{MAGENTA}{cmyk}{0,1,0,0}\definecolor{YELLOW}{cmyk}{0,0,1,0}}

\usepackage{dcolumn}
\usepackage{bm}
\usepackage{times}
\usepackage{hyperref}
\hypersetup{colorlinks=true,linkcolor=red,citecolor=blue}
\makeatother

\usepackage{babel}

\makeatother

\usepackage{babel}
\begin{document}

\title{Spiral magnetic ordering of the Eu moments in EuNi$_{2}$As$_{2}$}

\author{W. T. Jin}

\email{jwt2006@gmail.com}

\affiliation{Jülich Centre for Neutron Science JCNS at Heinz Maier-Leibnitz Zentrum (MLZ), Forschungszentrum Jülich GmbH, Lichtenbergstraße 1, D-85747 Garching, Germany}

\author{N. Qureshi}

\affiliation{Institut Laue Langevin, 71 rue des Martyrs, BP156, 38042 Grenoble Cedex 9, France}

\author{Z. Bukowski}

\affiliation{Institute of Low Temperature and Structure Research, Polish Academy of Sciences, 50-422 Wroclaw, Poland}

\author{Y. Xiao}

\affiliation{Jülich Centre for Neutron Science JCNS and Peter Grünberg Institut PGI, JARA-FIT, Forschungszentrum Jülich GmbH, D-52425 Jülich, Germany}

\author{S. Nandi}

\affiliation{Department of Physics, Indian Institute of Technology, Kanpur 208016, India}

\author{M. Babij}

\affiliation{Institute of Low Temperature and Structure Research, Polish Academy of Sciences, 50-422 Wroclaw, Poland}

\author{Z. Fu}

\affiliation{Jülich Centre for Neutron Science JCNS at Heinz Maier-Leibnitz Zentrum (MLZ), Forschungszentrum Jülich GmbH, Lichtenbergstraße 1, D-85747 Garching, Germany}

\author{Y. Su}

\affiliation{Jülich Centre for Neutron Science JCNS at Heinz Maier-Leibnitz Zentrum (MLZ), Forschungszentrum Jülich GmbH, Lichtenbergstraße 1, D-85747 Garching, Germany}

\author{Th. Brückel}

\affiliation{Jülich Centre for Neutron Science JCNS and Peter Grünberg Institut PGI, JARA-FIT, Forschungszentrum Jülich GmbH, D-52425 Jülich, Germany}

\affiliation{Jülich Centre for Neutron Science JCNS at Heinz Maier-Leibnitz Zentrum (MLZ), Forschungszentrum Jülich GmbH, Lichtenbergstraße 1, D-85747 Garching, Germany}

\begin{abstract}
The ground-state magnetic structure of EuNi$_{2}$As$_{2}$ was investigated by single-crystal neutron diffraction. At base temperature, the Eu$^{2+}$ moments are found to form an incommensurate antiferromagnetic spiral-like structure with a magnetic propagation vector of $\mathit{k}$ = (0, 0, 0.92). They align ferromagnetically in the $\mathit{ab}$ plane with the moment size of 6.75(6) $\mu_{B}$, but rotate spirally by 165.6(1)° around the $\mathit{c}$ axis from layer to layer. The magnetic order parameter in the critical region close to the ordering temperature, $\mathit{T_{N}}$ = 15 K, shows critical behavior with a critical exponent of $\beta_{Eu}$ = 0.34(1), consistent with the three-dimensional Heisenberg model. Moreover, within the experimental uncertainty, our neutron data is consistent with a model in which the Ni sublattice is not magnetically ordered. 
\end{abstract}

\maketitle

\section{Introduction}

Since the discovery of unconventional superconductvity (SC) in iron pnictides in 2008,\cite{Kamihara_08} the ternary ``122'' compounds $\mathit{A}$Fe$_{2}$As$_{2}$ (with $\mathit{A}$ = Ba, Sr, Ca or Eu) crystallizing in the tetragonal ThCr$_{2}$Si$_{2}$ type structure are drawing persistent research interest, due to the high superconducting transition temperature ($\mathit{T_{SC}}$) up to 38 K achievable in these materials.\cite{Rotter_08,Sefat_08} Among them, EuFe$_{2}$As$_{2}$ is a unique member of the ``122'' family, since the $\mathit{A}$ site is occupied by the $\mathit{S}$-state rare earth Eu$^{2+}$ ion possessing a 4$f^{7}$ electronic configuration with the electron spin $\mathit{S}$ = 7/2.\cite{Marchand_78} Upon chemical doping or applying external pressure, EuFe$_{2}$As$_{2}$ can be tuned into a superconductor with $\mathit{T_{SC}}$ up to 32 K.\cite{Jeevan_08,Ren_09,Jiao_11,Jiao_13,Jin_PhaseDiagram,Miclea_09,Terashima_09}

During the past few years, many experimental efforts have been devoted to studying the intriguing magnetism in the EuFe$_{2}$As$_{2}$-based compounds and its interplay with the SC. The parent compound, EuFe$_{2}$As$_{2}$, displays a spin-density-wave (SDW) ordering of the itinerant Fe$^{2+}$ moments concomitant with a tetragonal-to-orthorhombic structural phase transition below 190 K. In addition, the localized Eu$^{2+}$ spins order below 19 K in an A-type antiferromagnetic (AFM) structure, in which the ferromagnetic (FM) layers with moments along the orthorhombic $\mathit{a}$ axis stack antiferromagnetically along the $\mathit{c}$ axis.\cite{Jiang_09_NJP,Xiao_09,Herrero-Martin_09} This magnetic configuration of Eu is stable with respect to chemical substitutions at small doping levels,\cite{Jeevan_11,Jin_Polarization} but the Eu$^{2+}$ spins start to cant out of the $\mathit{ab}$ plane at intermediate doping levels\cite{Guguchia_11_torque,Jin_Pressure,Jin_PhaseDiagram,Blachowski_11} and become ferromagnetically aligned along the $\mathit{c}$ axis
at relatively high doping levels.\cite{Jin_13,Nandi_14_neutron,Nandi_14,Jin_15} Interestingly, the ferromagnetism from the Eu sublattice was found to be compatible with the SC induced by either chemical doping or hydrostatic pressure,\cite{Nandi_14,Jin_Pressure,Jin_PhaseDiagram} making the ``ferromagnetic superconductors'' in the EuFe$_{2}$As$_{2}$-based system very attractive research subjects.

Recently, bulk superconductivity has been discovered in the Ni-based ``122'' pnictides, Ba(Ni$_{1-x}$Co$_{x}$)$_{2}$As$_{2}$.\cite{Eckberg_18} This motivated us to revisit another member of the ``122'' nickel arsenides, EuNi$_{2}$As$_{2}$, a structural analogue of the well-studied EuFe$_{2}$As$_{2}$, as a better understanding of the properties of the parent compound is fundemental for realizing SC in this system by chemical doping or external pressure. Previously, the structural and physical properties of EuNi$_{2}$As$_{2}$ were investigated by different experimental methods. Using single-crystal x-ray diffraction, EuNi$_{2}$As$_{2}$ was found to possess a tetragonal symmetry (space group $\mathit{I4/mmm}$) at room temperature, similar to EuFe$_{2}$As$_{2}$.\cite{Jeitschko_88} The resistivity of EuNi$_{2}$As$_{2}$ shows a metallic behavior. It is almost linear within a wide temperature range from 30 K to 300 K, and a kink shows up at 14 K, due to the magnetic ordering of Eu.\cite{Ghadraoui_88,Bauer_08} The absence of any other anomalies in the temperature dependence of the resistivity is in stark contrast to that of EuFe$_{2}$As$_{2}$,\cite{Xiao_12} suggesting the absence of Fermi surface nesting due to structural distortions in EuNi$_{2}$As$_{2}$. In addition, magnetic susceptibility measurements on both polycrystalline and single-crystal samples of EuNi$_{2}$As$_{2}$ suggest an antiferromagnetic ordering of Eu below 14 K. However, to the best of our knowledge, no neutron diffraction experiments have been performed so far and the detailed magnetic structure of the Eu$^{2+}$ moments is unknown for EuNi$_{2}$As$_{2}$. 

Here we report the experimental results on the ground-state magnetic structure and critical behavior of EuNi$_{2}$As$_{2}$, obtained through single-crystal neutron diffraction.

\section{Experimental Details}

Platelet-like single crystals of EuNi$_{2}$As$_{2}$ were flux grown, using Bi as flux. X-ray Laue diffraction confirmed that the crystals have the c axis perpendicular to their surfaces. A 22 mg single crystal with dimensions $\backsim$ 4 × 4 × 0.3 mm$^{3}$ was selected for
the neutron diffraction measurement. It was performed on the 4-circle diffractometer D10 at the Institut Laue–Langevin (Grenoble, France) using the neutron wavelength of 2.36 Å. A pyrolytic graphite filter was employed to suppress the higher harmonics of the primary beam
to less than 10$^{\lyxmathsym{–}4}$. The crystal was mounted on top of an aluminum pin with a small amount of GE varnish and put inside a closed-cycle cryostat in the 4-circle geometry. The mosaic width of $\backsim$ 0.3° in the rocking scans confirms the good bulk quality of the crystal. For collecting the integrated intensities of accessible nuclear and magnetic reflections, the two-dimensional (2D) area detector was used. For reciprocal-space Q scans, a flat pyrolytic graphite analyzer was used to reduce the background. A 2.2 mg crystal from
the same batch was chosen for the magnetization measurements using the Quantum Design magnetic property measurement system (MPMS).

\section{Experimental Results }

Figure 1(a) shows the temperature dependencies of the dc magnetic susceptibility ($\chi$) of the EuNi$_{2}$As$_{2}$ single crystal, measured under the zero-field-cooling (ZFC) and field-cooling (FC) conditions in an applied magnetic field of 100 Oe parallel to the $\mathit{ab}$ plane and $\mathit{c}$-axis, respectively. Above 20 K, $\chi_{ab}$ and $\chi_{c}$ almost overlap, indicating an isotropic susceptibility. Below 14.5 K, however, a significant anisotropy shows up. $\chi_{ab}$ drops sharply, while $\chi_{c}$ remains almost constant with decreasing temperature, suggesting an antiferromagnetic (AFM) alignment of the Eu$^{2+}$ moments in the $\mathit{ab}$ plane. The Neel temperature of $\mathit{T_{N}}$ = 14.5(5) K is consistent with a similar value of 14 K from previous magnetic susceptibility measurements on polycrystalline and single-crystal EuNi$_{2}$As$_{2}$.\cite{Raffius_93,Ghadraoui_88} Fig. 1(b) shows the in-plane magnetic susceptibility ($\chi_{ab}$) measured in the field of 0.1 T. By fitting the paramagnetic region (from 50 K to 200 K) with the Curie-Weiss law, $\chi^{-1}$ = $\frac{T-\theta}{C}$, the Curie constant and Weiss temperature can be deduced to be $\mathit{C}$ = 8.50(4) emu K mol$^{-1}$ and $\theta$ = -17.7(8) K, respectively. The negative value of $\theta$ indicates an AFM interaction between the Eu$^{2+}$ spins. The effective paramagnetic moment of Eu is estimated to be $\mu_{eff}$ = 8.22(3) $\mu_{B}$, close to the theoretical value of $\mathit{g\sqrt{S(S+1)}}$ = 7.94 $\mu_{B}$ with $\mathit{S}$ = $\frac{7}{2}$ and the Landé factor$\mathbf{\mathit{g}}$ = 2. 

\begin{figure}
\centering{}\includegraphics{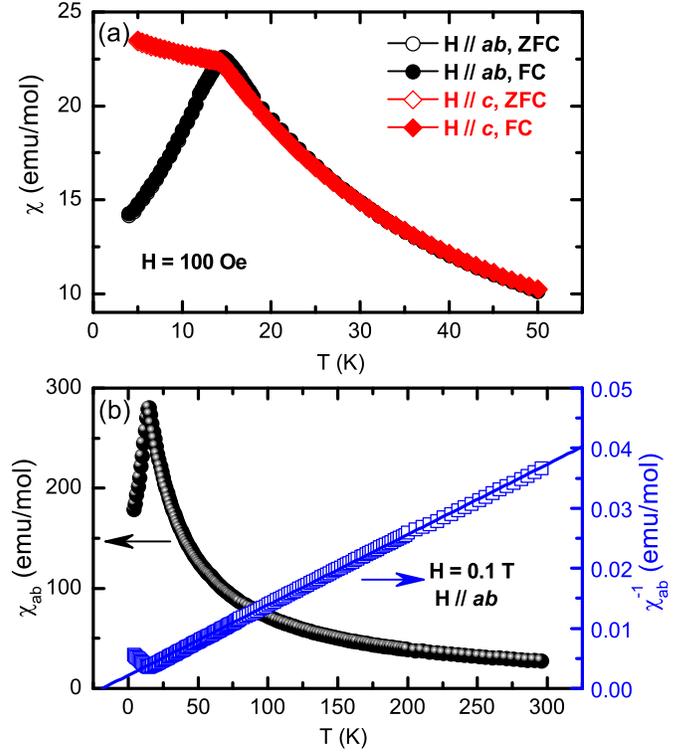}
\caption{(a) The temperature dependencies of the dc magnetic susceptibility of the EuNi$_{2}$As$_{2}$ single crystal, measured under the ZFC and FC conditions in an applied field of 100 Oe parallel to the $\mathit{ab}$ plane and $\mathit{c}$-axis, respectively. No history dependence is evidenced, i.e., FC and ZFC curves are almost identical. (b) The in-plane magnetic susceptibility measured in the field of 0.1 T, and the fitting to the paramagnetic region of inverse susceptibility using the Curie-Weiss law (solid line).}
\end{figure}

To clarify the magnetic ground state of EuNi$_{2}$As$_{2}$ in detail, single-crystal neutron diffraction experiment was carried out as a microscopic probing method. First of all, a set of one-dimensional and two-dimensional reciprocal-space scans were performed at the base
temperature to search for the magnetic reflections from the Eu sublattice, see Fig. 2 and 3, respectively. As shown in Fig.2, the scans at 2 K along (0 0 $\mathit{L}$), (1 0 $\mathit{L}$) and (1 1 $\mathit{L}$) directions do not show additional intensities on top of the nuclear
reflections with $\mathit{k}$ = (0, 0, 0), compared with those at 20 K (the tetragonal notation is used throughout this paper). The magnetic scattering intensities also do not emerge at the forbidden Bragg peak positions with the propagation vector $\mathit{k}$ = (0, 0, 1). The FM ordering and A-type AFM ordering of the Eu$^{2+}$ moments as reported previously for EuCr$_{2}$As$_{2}$ and EuFe$_{2}$As$_{2}$ can therefore be excluded for EuNi$_{2}$As$_{2}$. Instead, a set of satellite magnetic peaks show up at both sides of the nuclear reflections, with the propagation vector $\mathit{k}$ = $\pm$ (0, 0, 0.92). Moreover, in addition to the satellite reflections at (0, 0, even $\pm$ 0.92), (1, 1, even $\pm$ 0.92) and (1, 0, odd $\pm$ 0.92), no additional peaks are observed at other Q positions in the ($\mathit{H}$, $\mathit{H},$ $\mathit{L}$) (Fig. 3a and 3b) and ($\mathit{H}$, $\mathit{0},$ $\mathit{L}$) (Fig. 3c and 3d) planes, as shown in the two-dimensional mesh scans. Based on these results, we could reach the conclusion that the Eu$^{2+}$ moments in EuNi$_{2}$As$_{2}$ are ordered in a single-$\mathit{k}$ incommensurate magnetic structure. 

\begin{figure}
\centering{}\includegraphics{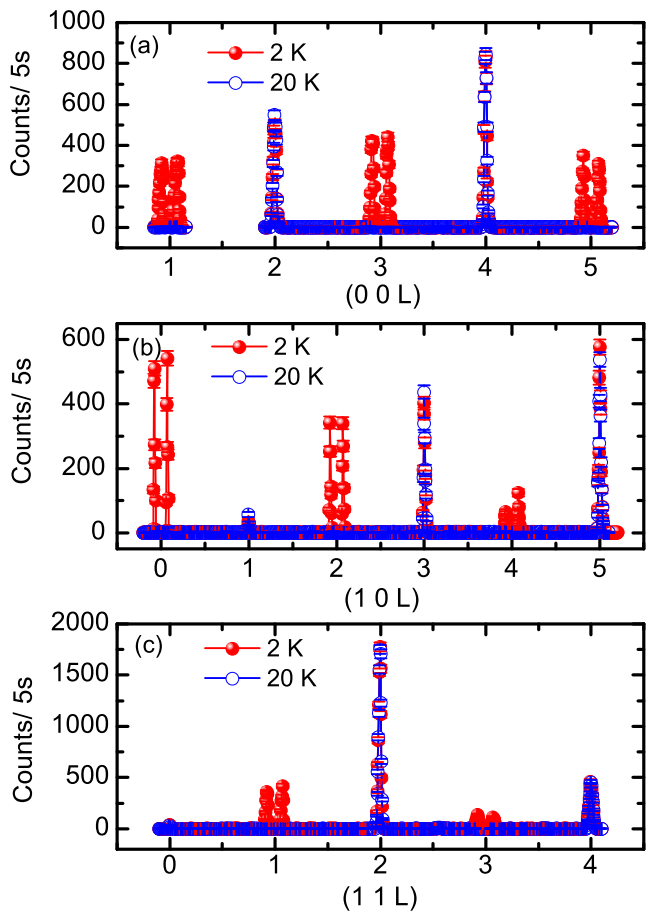}
\caption{Reciprocal-space scans along the (0 0 $\mathit{L}$) (a), (1 0 $\mathit{L}$) (b), and (1 1 $\mathit{L}$) (c) directions at 2 K and 20 K, respectively.} 
\end{figure}

\begin{figure*}
\centering{}\includegraphics[width=1\textwidth]{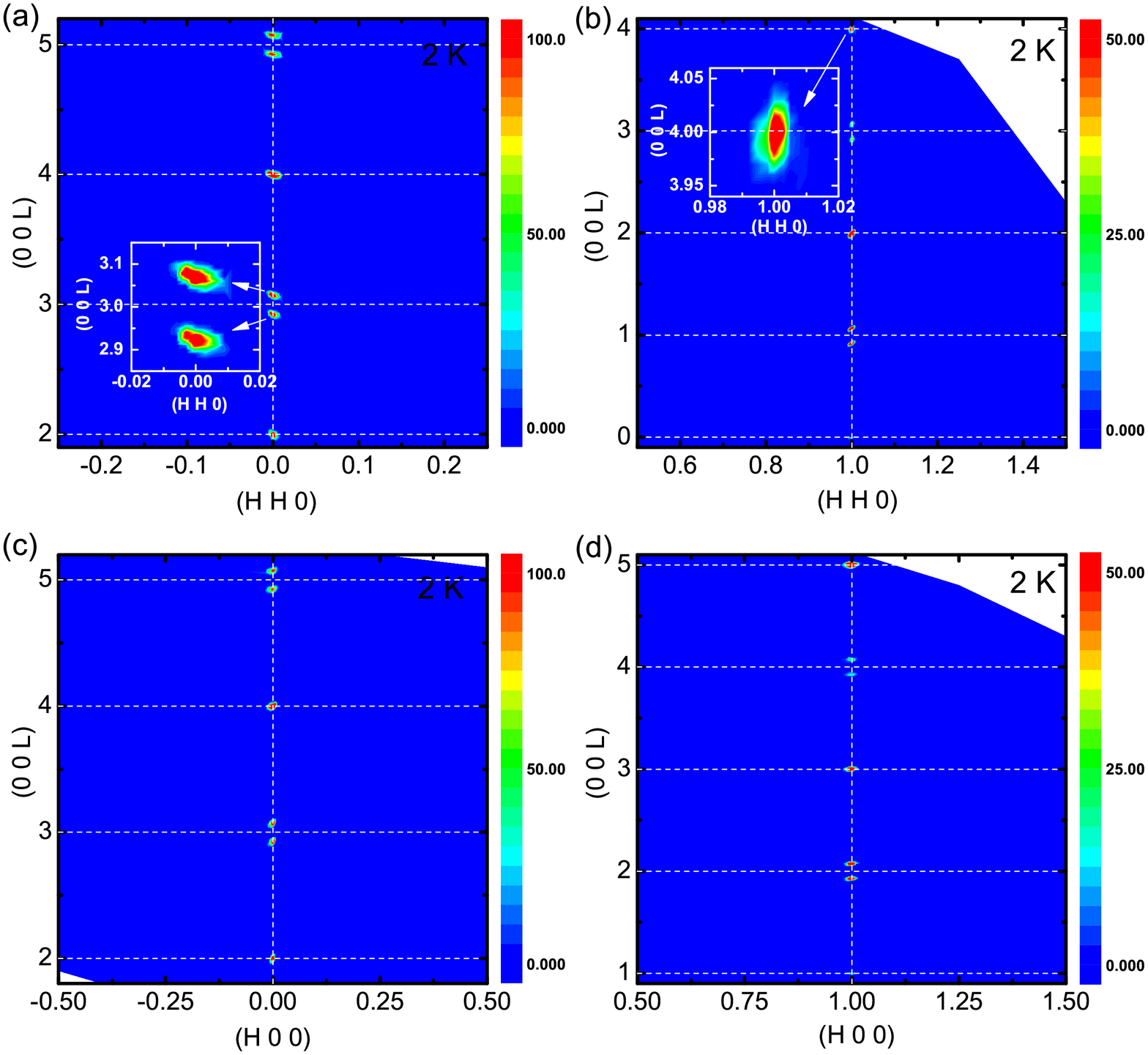}
\caption{Two-dimensional mesh scans at 2 K in ($\mathit{H}$, $\mathit{H},$ $\mathit{L}$) (a, b) and ($\mathit{H}$, $\mathit{0},$ $\mathit{L}$) (c, d) plane, indicating a magnetic propagation vector $\mathit{k}$ = (0, 0, 0.92) for EuNi$_{2}$As$_{2}$. Insets in (a) and (b) show the enlarged views of the magnetic satellite peaks around (0, 0, 3) and the non-split nuclear peak (1, 1, 4), respectively.}
\end{figure*}

The temperature dependence of the integrated intensity of the incommensurate magnetic reflection, (0, 0, 2.92), is plotted in Fig. 4, from which the AFM ordering temperature was estimated to be 15 K. As illustrated by the scans along the $\mathit{L}$ direction in the left inset,
the magnetic scattering intensity completely disappears at 20 K. Due to the small size of the single crystal, we can not observe the magnetic diffuse scattering due to spin fluctuations above the transition temperature. In order to determine the critical behaviour close to the phase transition, we have measured dense data points in the temperature range between 12 K and 15 K, utilizing the excellent stability of temperature control at D10. Within this critical region, the magnetic order parameter $\mathit{M}$ was fitted using the mean-field model, $\mathit{I}$ $\propto$ $M^{2}$ $\propto$ $\tau^{2\beta}$ , where $\tau$ = $\frac{T_{N}-T}{T_{N}}$. The right inset of Fig. 4 shows the linear fitting (dashed line) of $\mathit{I(\tau)}$ in the double logarithmic plot, from which the transition temperature and the critical exponent were deduced to be $\mathit{T_{N}}$ = 14.98(1) K and $\beta_{Eu}$ = 0.34(1), respectively. The $\mathit{T_{N}}$ value obtained here is consistent with $\mathit{T_{N}}$ = 14.5(5) K from the magnetic susceptibility data above. The critical exponent extracted here is close to the values obtained for other parent compouds of Eu-based pnictides, i.e., 0.350(8) for the A-type AFM ordering in EuFe$_{2}$As$_{2}$ and 0.32(2) for the incommensurate AFM ordering in EuRh$_{2}$As$_{2}$,\cite{Koo_10,Nandi_09} respectively. All are well consistent with the three-dimensional (3D) Heisenberg model ($\beta$ = 0.36).

\begin{figure}
\centering{}\includegraphics{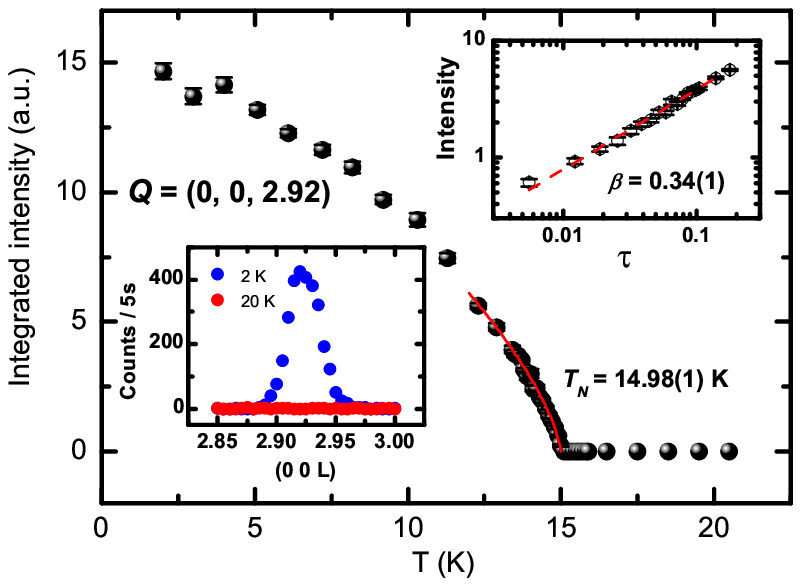}
\caption{The temperature dependence of the integrated intensity of the incommensurate magnetic reflection, (0, 0, 2.92), in which the solid line represents the fitting using the mean-field model in the critical region. The scans along the $\mathit{L}$ direction are shown in the left inset, for 2 K and 20 K, respectively. The right inset shows the linear fitting (dashed line) of $\mathit{I(\tau)}$ in the double logarithmic plot. }
\end{figure}

In order to determine the detailed magnetic structure of the Eu$^{2+}$ moments in EuNi$_{2}$As$_{2}$, the integrated intensities of 132 nuclear reflections allowed by the space group symmetry $\mathit{I4/mmm}$ and 119 satellite magnetic reflections were collected using the 2D detector at 2K. After necessary absorption correction procedure using the DATAP program by inputing the dimensions of the crystal,\cite{Coppens_65} the equivalent reflections were merged into the unique ones based on the tetragonal symmetry. The nuclear structure was firstly refined using the FULLPROF program within the $\mathit{I4/mmm}$ space group,\cite{Rodriguez-Carvajal_93} since no evidence for a tetragonal-to-orthorhombic structural phase transition was found in our neutron measurements, consistent with previous resistivity data.\cite{Ghadraoui_88,Bauer_08} As shown in the inset of Fig. 3(b), the (1, 1, 4) reflection does not exhibit any splitting at 2 K within the experimental resolution, supporting the tetragonal symmetry. The results of the refinements are listed in Table I. Considering the difficulty associated with the absorption correction on our irregular-shaped crystal, the calculated and observed intensities of the nonequivalent nuclear reflections as shown in Fig. 5(a) are in good agreement. 

\begin{table}
\caption{Parameters for the nuclear and magnetic structures of EuNi\textsubscript{2}As\textsubscript{2 }at 2 K obtained from refinements of single-crystal neutron diffraction data. The isotropic temperature factors ($\mathit{B}$) of all atoms were refined. (Space group: $\mathit{I4/mmm}$, $\mathit{a}$ = 4.1015(3) Å, $\mathit{c}$ = 10.0466(5) Å)}
\begin{ruledtabular} %
\begin{tabular}{ccccc}
\hline 
Atom/site & $\mathit{x}$ & $\mathit{y}$ & $\mathsf{\mathit{z}}$ & $B\,$(Å\textsuperscript{2})\tabularnewline
\hline 
Eu ($2a$) & 0 & 0 & 0 & 0.7(1)\tabularnewline
\multicolumn{5}{c}{ $\mu_{Eu}$= 6.75(6) $\mu_{B}$, $\phi$ = 165.6(1)°}\tabularnewline
Ni ($4d$) & 0.5 & 0 & 0.25 & 0.4(1)\tabularnewline
As ($4e$) & 0 & 0  & 0.3683(6)  & 0.2(1)\tabularnewline
\hline 
\multicolumn{5}{c}{Extinction g (rad$^{\lyxmathsym{−}1}$) = 2.1(9)}\tabularnewline
\multicolumn{5}{c}{$R{}_{F^{2}}$, $R{}_{wF^{2}}$, $R_{F}$ , $\chi^{2}$ (nuclear):
11.3, 8.9, 6.21, 6.47 }\tabularnewline
\multicolumn{5}{c}{$R{}_{F^{2}}$, $R{}_{wF^{2}}$, $R_{F}$ , $\chi^{2}$ (magnetic):
7.48, 9.61, 3.85, 9.43 }\tabularnewline
\end{tabular}\end{ruledtabular} 
\end{table}

According to the representation analysis,\cite{Wills_00} for the space group of $\mathit{I4/mmm}$, only two magnetic representations are possible for the Eu ($2a$) site with the propagation vector of $\mathit{k}$ = (0, 0, 0.92) , which we label as $\Gamma_{1}$ and
$\Gamma_{5}$, respectively. $\Gamma_{1}$ allows the $\mathit{c}$-axis aligned ferromagnetic Eu layers stacking antiferromagnetically, with varying moment size at differnt layers. This model is not consistent with the magnetization data as shown in Fig. 1(a), and also gives
a very poor fit to the magnetic intensities. On the other hand, $\Gamma_{5}$ allows the in-plane aligned ferromagnetic Eu layers stacking spirally along the $\mathit{c}$ axis, with a constant moment size at differant layers. This model fits pretty well with the magnetic intensities, as shown in Table 2 and Fig. 5(b). The moment size of Eu is refined to be 6.75(6) $\mu_{B}$. As illustrated in Fig. 5(c), the Eu$^{2+}$ moments form an incommensurate spiral-like structure, with the moment direction lying in the $\mathit{ab}$ plane but rotating spirally by 165.6(1)° around the $\mathit{c}$ axis with respect to adjacent Eu layers. This magnetic configuration displays an overall AFM character, which agrees well with the magnetic susceptibility data shown in Fig. 1.

\begin{table}
\caption{Comparison between the observed intensities of the nonequivalent magnetic reflections at 2 K and calculated intensities using the $\Gamma_{5}$ magnetic structure model. The numbers in the parentheses correspond to the error bars of the observed intensity.}
\begin{ruledtabular} %
\begin{tabular}{ccccc|ccccc}
$\mathit{H}$ & $K$ & $\mathsf{\mathit{L}}$ & $\mathit{I_{obs}}$ & $\mathit{I_{cal}}$ & $\mathit{H}$ & $K$ & $\mathsf{\mathit{L}}$ & $\mathit{I_{obs}}$ & $\mathit{I_{cal}}$\tabularnewline
\hline 
0 & 0 & 1.08 & 3517(65) & 3472 & 1 & 1 & 1.08 & 1393(66) & 1461\tabularnewline
0 & 0 & 2.92 & 2794(44) & 3045 & 1 & 1 & 2.92 & 1358(122) & 1455\tabularnewline
0 & 0  & 3.08 & 2942(44) & 2952 & 1 & 1 & 3.08 & 1516(214) & 1433\tabularnewline
0 & 0 & 4.92 & 1997(42) & 1794 & 1 & 1 & 4.92 & 1240(171) & 1014\tabularnewline
0 & 0 & 5.08 & 2004(41) & 1701 & 1 & 1 & 5.08 & 1260(44) & 971\tabularnewline
1 & 0 & 0.08 & 1737(33) & 1737 & 2 & 0 & 0.92 & 910(28) & 923\tabularnewline
1 & 0 & 1.92 & 1765(59) & 2074 & 2 & 0 & 1.08 & 924(51) & 924\tabularnewline
1 & 0 & 2.08 & 1989(149) & 2081 & 2 & 0 & 2.92 & 878(40) & 869\tabularnewline
1 & 0 & 3.92 & 1614(227) & 1707 & 2 & 0 & 3.08 & 844(80) & 855\tabularnewline
1 & 0 & 4.08 & 1416(107) & 1649 & 2 & 1 & 0.08 & 760(30) & 751\tabularnewline
1 & 0 & 5.92 & 1223(52) & 964 & 2 & 1 & 1.92 & 757(46) & 736\tabularnewline
1 & 1 & 0.92 & 1349(66) & 1448 & 2 & 1 & 2.08 & 697(46) & 732\tabularnewline
\end{tabular}\end{ruledtabular} 
\end{table}

\begin{figure}
\centering{}\includegraphics{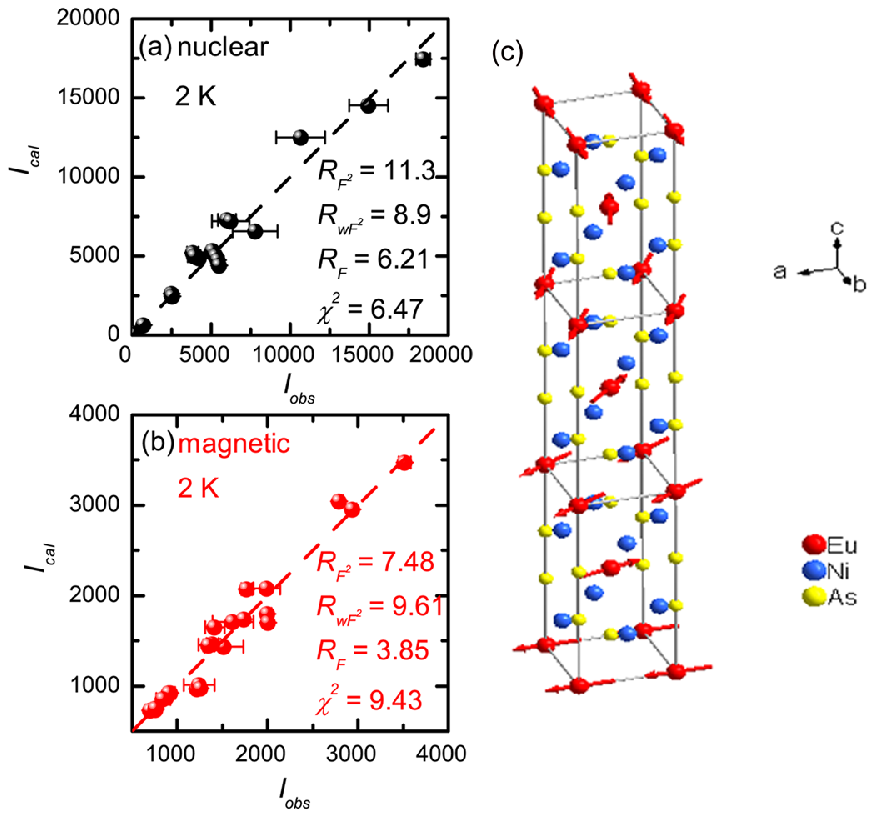}
\caption{Comparison between the observed and calculated integrated intensities of the nonequivalent nuclear (a) and magnetic (b) reflections, respectively, at 2 K, as well as the spiral-like magnetic structure (c) of EuNi\textsubscript{2}As$_{2}$ as determined by the refinements. }
\end{figure}

As revealed by neutron-diffraction measurements, in the ground state of EuFe\textsubscript{2}As$_{2}$ or EuCr$_{2}$As$_{2}$, the Fe or Cr sublattice is also magnetically ordered with the moment size of 0.98(8) $\mu_{B}$ and 1.7(4) $\mu_{B}$, respectively.\cite{Xiao_09,Nandi_16} However, in the case of EuCo\textsubscript{2}As$_{2}$, the moment size on the Co site was refined to be zero within the experimental uncertainty.\cite{Tan_16} In our case, the magnetic propagation vector of the Ni$^{2+}$ moments has to be $\mathit{k}$ = (0, 0, 0), if they are ordered, since no additional reflections were observed, in addition to the nuclear peaks and incommensurate magnetic peaks from Eu (see Fig. 3). Similar to EuCr$_{2}$As$_{2}$, four magnetic representations are possible for the $4d$ site according to the representation analysis, labeled as $\Gamma_{3}$, $\Gamma_{6}$ , $\Gamma_{9}$ and $\Gamma_{10}$, corresponding to FM alignment of the Ni$^{2+}$ moments along $\mathit{c}$ axis, AFM alignment along $\mathit{c}$ axis, FM alignment in the
$\mathit{ab}$ plane, and AFM alignment in the $\mathit{ab}$ plane, respectively. Adding these magnetic phases of Ni to the nuclear moel at 2 K does not yield visible improvement of the fitting, indicating that the moment size of Ni, if any, must be pretty small and beyond
the uncertainty of our experiment. Combined with the available macroscopic measurements performed on EuNi$_{2}$As$_{2}$ so far,\cite{Ghadraoui_88,Raffius_93,Bauer_08} from which no evidence of Ni magnetic ordering was found, we tentatively conclude that the Ni sublattice in EuNi$_{2}$As$_{2}$ is probably not magnetically ordered.

\section{Discussion and Conclusion}

The incommensurate AFM spiral-like structure in EuNi$_{2}$As$_{2}$ is distinct from the commensurate AFM or FM structure observed in EuFe$_{2}$As$_{2}$-based family and EuCr$_{2}$As$_{2}$,\cite{Xiao_09,Nandi_14,Nandi_14_neutron,Jin_15,Jin_PhaseDiagram,Nandi_16} but it is similar to that observed recently in EuCo$_{2}$As$_{2}$ with a different propagation vector of $\mathit{k}$ = (0, 0, 0.79).\cite{Tan_16} We note that such an in-plane spiral-like magnetic configuration is not unique to the ``122'' arsenides, as it was also reported for the ``122'' phosphides and germanides, EuCo$_{2}$P$_{2}$ and HoNi$_{2}$Ge$_{2}$, respectively.\cite{Raffius_93,Pinto_85} As is well recognized, the magnetic ordering of localized 4$\mathit{f}$ moments in the ``$\mathit{R}$$\mathit{M_{2}}X_{2}$'' compounds (where $\mathit{R}$ is rare-earth element and $\mathit{M}$ is a 3$\mathit{d}$ transition metal element) possessing the ThCr$_{2}$Si$_{2}$ structure arises from the indirect Ruderman-Kittel-Kasuya-Yosida (RKKY) interaction mediated by the conduction $\mathit{d}$ electrons.\cite{Akbari_13} Therefore, it is very likely that the strong dependence of the Eu magnetic configuration on the 3$\mathit{d}$ element $\mathit{M}$
($\mathit{M}$ = Fe, Co, Ni, Cr) in Eu$M$$_{2}$As$_{2}$ is due to the change of band structure of the conduction electrons contributed by different 3$\mathit{d}$ transition metals. Future theoretical studies on EuNi$_{2}$As$_{2}$ will be crucial for understanding
the origin of its intriguing incommensurate magnetic structure of the Eu sublattice, as well as clarifying the absence or presence of magnetic ordering of the Ni sublattice.

In conclusion, the ground-state magnetic structure of EuNi$_{2}$As$_{2}$ was investigated by single-crystal neutron diffraction. At base temperature, the Eu$^{2+}$ moments are found to form an incommensurate AFM spiral-like structure with a magnetic propagation vector of $\mathit{k}$ = (0, 0, 0.92). They align ferromagnetically in the $\mathit{ab}$ plane with the moment size of 6.75(6) $\mu_{B}$, but rotate spirally by 165.6(1)° around the $\mathit{c}$ axis from layer to layer. In addition, the critical behavior of the Eu magnetic ordering was studied. By fitting the magnetic order parameter in the critical region close to the ordering temperature, $\mathit{T_{N}}$ = 15 K, a critical exponent of $\beta_{Eu}$ = 0.34(1) is extracted, well consistent with that of the three-dimensional Heisenberg model. Moreover, within
the experimental uncertainty, our neutron data is consistent with a model in which the Ni sublattice is not magnetically ordered. 

\bibliographystyle{apsrev} \bibliographystyle{apsrev}
\begin{acknowledgments}
W. T. J. would like to acknowledge S. Mayr for the assistance with the orientation of the crystal, and Shang Gao for fruitful discussions. 
\bibliographystyle{apsrev}
\end{acknowledgments}

\end{document}